\documentclass[12pt]{iopart}
\input epsf 
\usepackage{iopams}
\usepackage[dvips]{graphicx}

\newtheorem{theorem}{Theorem}

\begin{document}
\title[Synchronised Similar Triangles for Three-Body Orbit with $L=0$]
{Synchronised Similar Triangles 
for Three-Body Orbit
with Zero Angular Momentum
}
\author{
Toshiaki~Fujiwara\dag,
Hiroshi~Fukuda\ddag,
Atsushi~Kameyama$\S$,
Hiroshi~Ozaki$\P$
and Michio~Yamada{*}
}
\address{\dag\ College of Liberal Arts and Sciences, Kitasato University, 
1-15-1 Kitasato, Sagamihara, Kanagawa 228-8555, Japan}%
\address{\ddag\ School of Administration and Informatics,
University of Shizuoka, 
52-1 Yada, Shizuoka 422-8526, Japan}
\address{\S\ Department of Mathematical and Design Engineering,
Faculty of Engineering, Gifu University,
1-1 Yanagido, Gifu, Gifu 501-1193, Japan
}
\address{\P\ Department of Physics, Tokai University,
1117 Kitakaname, Hiratsuka, Kanagawa 259-1292, Japan}
\address{{*}\ 
Research Institute for Mathematical Sciences,
Kyoto University,
Oiwake-cho, Kitashirakawa, Sakyo-ku, Kyoto 606-8502, Japan
}

\eads{
	\mailto{\dag\ fujiwara@clas.kitasato-u.ac.jp},
	\mailto{\ddag\ fukuda@u-shizuoka-ken.ac.jp},
	\mailto{\S\ kameyama@cc.gifu-u.ac.jp},
	\mailto{\P\ ozaki@keyaki.cc.u-tokai.ac.jp}, 
	\mailto{{*}~yamada@kurims.kyoto-u.ac.jp}
}

\begin{abstract}
Geometrical properties of
three-body orbits with
zero angular momentum
are investigated.

If the moment of inertia is also
constant along the orbit,
the triangle whose vertexes are the positions of the bodies,
and
the triangle whose perimeters are the momenta of the bodies,
are always similar
(``synchronised similar triangles'').
This similarity yields
kinematic equalities 
between
mutual distances and magnitude of momenta.
Moreover,
if the orbit is a solution to
the equation of motion
under homogeneous potential,
the orbit has a new constant
involving momenta.

For orbits
with zero angular momentum
and non-constant moment of inertia,
we introduce
scaled variables,
positions divided by
square root of the moment of inertia
and momenta derived from
the velocity of the scaled positions.
Then the similarity and
the kinematic equalities
hold for the scaled variables.
Using this similarity,
we prove
that
any bounded three-body orbit
with zero angular momentum
under homogeneous potential
whose degree is smaller than 2
has 
infinitely many
collinear configurations
(syzygies or eclipses)
or collisions.
\end{abstract}


\section{Introduction and summary}
Recently,
the figure-eight solution to 
the planer equal-masses three-body problem 
was
found by
Moore, Chenciner \& Montgomery 
and Sim\'{o}~\cite{moore,chenAndMont,simo1,simo2},
and 
is paid attention to.
Numerical calculations shows that
this solution is unique 
up to 
translation,
rotation
and
scale transformation.
Therefore, we write this solution 
``the figure-eight''.

The figure-eight solution has zero angular momentum.
Fujiwara, Fukuda and Ozaki~(FFO)  \cite{ffoLem} pointed out that
this gives an important information of the orbit:
three tangent lines 
at 
the three bodies
meet at a point at each instant.
Below we 
call this theorem the ``three-tangents theorem'',
and the crossing point  the ``centre of tangents'' $C_t$.
FFO used this theorem to find 
a figure-eight solution on the lemniscate curve~\cite{ffoLem}.

Then natural questions arise.
What will happen
if three normal lines meet at a point?
What conditions make three normal lines meet at a point?
We 
show in this paper
that
in the planer three-body problem with general masses, 
if the moment of inertia is constant along the orbit,
then
three normal lines 
at
the bodies meet at a point.
We 
now
call this theorem the ``three normals theorem'',
and the crossing point the ``centre of normals'' $C_n$.


The first half of this paper
is devoted to clarify the nature
of orbits
with zero angular momentum
and constant moment of inertia.
These orbits have
the following two
geometrical properties. 
The first property is the following.
The centre of tangents $C_t$ and the centre of normals $C_n$
are the end points of a diameter
of the circumcircle for the triangle made by three bodies.
Thus the midpoint of $C_t$ and $C_n$
is the circumcenter $C_o$.
%
%
\begin{figure}[!h]
\begin{center}
\includegraphics[width=3in]{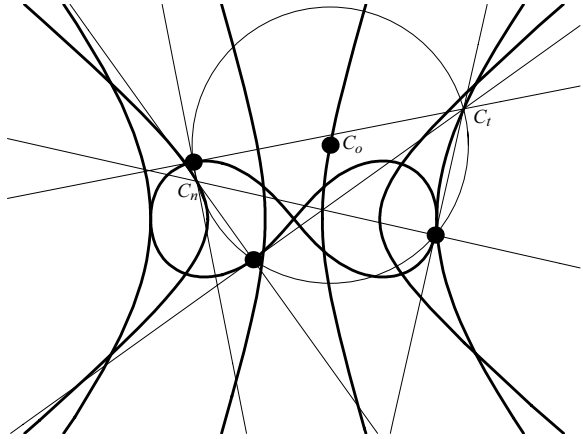}
\caption{The eight-shaped curve is 
the figure-eight orbit under the potential $-1/r_{ij}^2$.
Hyperbola-like curves from inner to outer are
the orbits of  $C_o$,
 $C_n$
and  $C_t$
respectively.
Tangent lines and normal lines
are represented by thin lines.
}
\label{figCtCnCo}
\end{center}
\end{figure}
In \fref{figCtCnCo},
the figure-eight solution
under the interaction potential
$-1/r_{ij}^2$
is shown,
where $r_{ij}$ is the distance between bodies
$i$ and $j$.
It is known that
the figure-eight solution
under this potential  has
zero angular momentum
and
a constant moment of inertia~\cite{chenq13}.
%
The positions of three bodies
and the circumcenter $C_o$ are
represented by solid circles.
The big circle is the circumcircle
and
we can find both the $C_t$ and $C_n$ on it,
though the diameter between $C_t$ and $C_n$
are not shown.

\begin{figure}[!h]
\begin{center}
\includegraphics[width=5in]{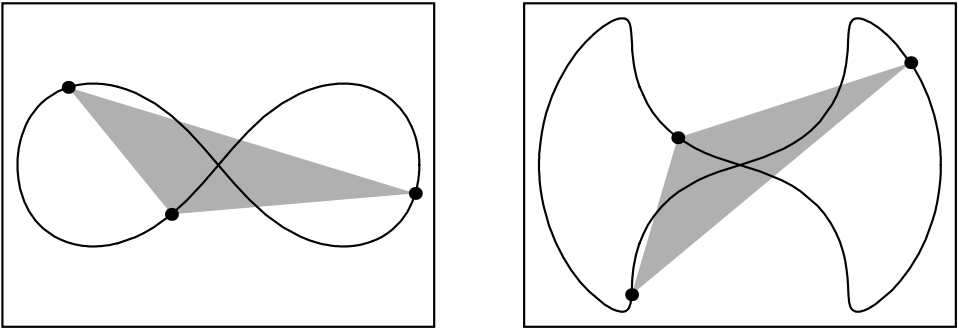}
\caption{
The figure-eight orbit under $-1/r_{ij}^2$ potential, with $m_i=1$.
Left: Orbit for $q_i^{\prime}=q_i/\sqrt{I}$.
Right: Orbit for
$p_k^{\prime}=(p_i-p_j)/\sqrt{3K}$,
where $(i, j, k)$ is a cyclic permutation of $(1,2,3)$.
These two triangles (grey areas) are always congruent
with each other
in inverse orientation.
}
\label{figXPforFigureEight}
\end{center}
\end{figure}

The second property is
a consequence of the first property.
The triangle
whose vertexes are the positions
$q_i$, 
and the triangle
whose perimeters are the momenta
$p_i$, 
are always inversely similar
(similar in inverse orientation).
In other words,
the triangle whose perimeters are 
the mutual distances $r_{ij}=|q_i-q_j|$, 
and the triangle whose perimeters are 
the magnitude of  the momenta $|p_k|$,
are always inversely similar.
See \fref{figXPforFigureEight}.
We 
call these triangles
the ``synchronised similar triangles'',
because they are always similar.

We will show 
in the following
that
this similarity yields the following
kinematic equalities
\begin{equation}
 \label{eqDefOfRatio0}
\frac{|p_1(t)|}{|q_2(t)-q_3(t)|}
 =\frac{|p_2(t)|}{|q_3(t)-q_1(t)|}
 =\frac{|p_3(t)|}{|q_1(t)-q_2(t)|}
 =\sqrt{
	\frac{m_1 m_2 m_3 K(t)}{M I}
	},
\end{equation}
\begin{equation}
\label{eqArea0}
\frac{q_i\wedge q_j}{I}
+
\frac{v_i\wedge v_j}{K}
=0
\end{equation}
where
$K$, $M$ and $I$ represent
twice of the kinetic energy, the total mass and 
the moment of inertia around the centre of mass,
respectively.
The symbol $\wedge$ represents outer product.
As shown in the equation \eref{eqDefOfRatio0},
the mutual distances $r_{ij}=|q_i-q_j|$ and 
the magnitude of the momenta $|p_k|$
are strongly related.

Moreover, we will show that
the solution orbit
to the
equation of motion
under the homogeneous potential
\begin{equation}
\label{eqDefOfV}
V_\alpha=
\cases{
\frac{1}{\alpha}
	\sum_{i<j}m_i m_j r_{ij}^\alpha	&for $\alpha\ne 0$\\
\sum_{i<j}m_i m_j \log r_{ij}		&for $\alpha=0$,
}
\end{equation}
has
a new constant along 
it:
\begin{equation}
  \label{eqConstant0}
  \cases{
  \sum_{(i, j, k)}m_i m_j |p_k|^\alpha=\mbox{constant}	&for $\alpha\ne 0$\\
  \sum_{(i,j,k)}m_i m_j \log |p_k|=\mbox{constant}		&for $\alpha=0$,
  }
\end{equation}
where
$(i,j,k)$ runs for the cyclic permutations of $(1,2,3)$.

%
%
The latter half 
of this paper is devoted 
to clarify the nature of
orbits
with zero angular momentum
and non-constant moment of inertia.
For these orbits, 
consider the following scaled variables
\begin{eqnarray}
  \label{eq_scaledq}
  \tilde{q}_i=\frac{q_i}{\sqrt{I}},\\
  \label{eq_scaledv}
  \tilde{v}_i=\frac{d \tilde{q}_i}{dt},
\end{eqnarray}
in order to make the scaled moment of inertia constant.
Then 
the triangle whose vertexes are $\tilde{q_i}$,
and 
the triangle whose perimeters are $m_k \tilde{v}_k$,
are the 
``synchronised similar triangles".
The kinematic equalities \eref{eqDefOfRatio0} and \eref{eqArea0}
hold for these scaled variables,
whereas
the equation \eref{eqConstant0} 
does not.
Using this similarity,
we will get the following interesting equation
\begin{equation}
 \label{eq_AreaGeneral}
 K q_i \wedge q_j
 +
 I v_i \wedge v_j
 =
 \frac{1}{2}
 \frac{d I}{dt}
 \frac{d}{dt}(q_i \wedge q_j).
\end{equation}
This equation holds for
any three-body orbit
with zero angular momentum.

Then we will show that
the oriented area
\begin{equation}
\Delta=\frac{1}{2}(q_2-q_1)\wedge(q_3-q_1),\label{orArea}
\end{equation}
satisfies the following equation
under the potential $V_{\alpha}$
\begin{equation}
\label{eqForDelta}
	I
	\frac{d}{dt}
		\left(
			\frac{1}{I}
			\frac{d \Delta}{dt}
		\right)
	=
	-
	\left(
		\frac{2K}{I}
		+
		\sum_{k\ell} (m_k+m_\ell)r_{k\ell}^{\alpha-2}
	\right)
	\Delta.
\label{deq_Delta}
\end{equation}
From this equation,
we can easily prove that
any bounded three-body orbit
with zero angular momentum
has infinitely many
collinear configurations
(syzygies or eclipses)
or collisions,
if $\alpha \le 2$.
This
marvellous
theorem was first
formulated and proved by
Montgomery~\cite{mont},
who derived
an equation 
for $\Delta/I$ 
similar to the equation \eref{eqForDelta}
with an elaborate calculation.

In section \ref{basicProperty}, 
we prove the four geometrical theorems,
the ``three tangents",  
the ``three normals",
the ``circumcircle"
and the ``synchronised similar triangles".
In the same section,
we also prove a
purely algebraic theorem
which is
a generalisation of
the theorem of ``synchronised similar triangles".
In section \ref{sec_homopot}, 
a new constant (\ref{eqConstant0})
is deduced
along the orbit
under a homogeneous potential. 
%
In section \ref{sec_force_space},
we point out that
there also exist
``synchronised similar triangles" in
the momentum space and the force space.
The scaled variables (\ref{eq_scaledq}) and
(\ref{eq_scaledv})
are introduced in section \ref{sec_triangle_gen}
and then another proof for 
Montgomery's ``infinitely many syzygies''~\cite{mont} 
is given in section \ref{sec_syzygies}.
Finally in section \ref{sec_final} 
we discuss 
some related problems.

\section{Three tangents, three normals, 
circumcircle and synchronised similar triangles}
\label{basicProperty}
We now show some
geometric and 
kinematic properties
of planer three-body orbits
with zero angular momentum
and constant moment of inertia.

\begin{theorem}[Three Tangents]
\label{ThreeTangents}
If 
both
the linear momentum and the angular momentum are zero,
three tangent lines 
at
the bodies meet at a point
or three tangent lines are parallel.
\end{theorem}
\textbf{Proof:}
Assume two tangent lines 
at the
bodies 1 and 2 meet at a point $C_t$.
Since $\sum_i p_i =0$ and $\sum_i q_i \wedge p_i=0$,
we have $\sum_i (q_i-C_t) \wedge p_i=0$.
By the assumption, we have $(q_1-C_t) \wedge p_1=(q_2-C_t) \wedge p_2=0$.
Thus we get $(q_3-C_t) \wedge p_3=0$.
That is, the tangent line 
at
the body 3 also passes through the point $C_t$.
Since $\sum_i p_i =0$, it is obvious
that
if two tangent lines are parallel,
the third line 
is
also 
parallel to the other two lines.
\begin{theorem}[Three Normals]
\label{Three Normals}
If the linear momentum is zero and the moment of inertia is constant,
three normal lines 
at
the bodies meet at a point
or three normal lines are parallel.
\end{theorem}
\textbf{Proof:}
Similar argument holds for 
$\sum_i p_i =0$ and $\sum_i q_i \cdot p_i=0$.
\begin{theorem}[Circumcircle]
\label{Circumcircle}
If the linear momentum is zero,
the angular momentum is zero
and moment of inertia is constant, 
then
the points $C_t$ and $C_n$ are the 
endpoints
of a diameter of 
the
circumcircle
of the triangle made of $q_1,q_2,q_3$.
\end{theorem}
\textbf{Proof:}
This is because
the angles $C_t$--$q_i$--$C_n$  are 90 degrees
for $i=1,2,3$.
\begin{theorem}[Synchronised Similar Triangles]
\label{SimilarTriangles}
If the linear momentum is zero,
the angular momentum is zero
and moment of inertia is constant,
then
the triangle whose vertexes are $q_i$
and the triangle whose perimeters are $p_i$,
are always 
inversely similar.
\end{theorem}
\begin{figure}[!h]
\begin{center}
\includegraphics[width=2in]{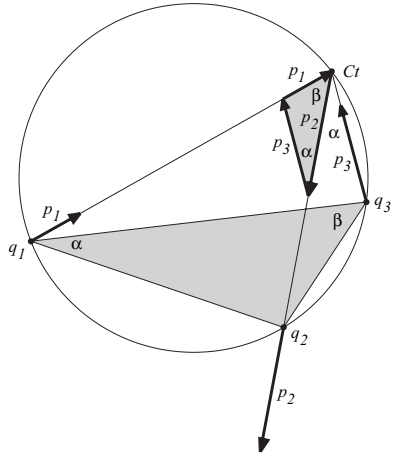}
\caption{
The triangle whose vertexes are $q_i$
(large grey triangle) 
and the triangle whose perimeters are $p_i$ 
(small grey triangle) are 
inversely similar.
}
\label{figXPtriangle4}
\end{center}
\end{figure}
\textbf{Proof:}
Since $\sum_i p_i =0$,
vectors $p_1$, $p_2$, $p_3$ form a triangle.
By the theorem \ref{Circumcircle},
points $q_1$, $q_2$, $q_3$ and $C_t$ are on the circumcircle.
Then,
the angles denoted by $\alpha$ in
\fref{figXPtriangle4} are identical.
The angles 
denoted by $\beta$ are also identical.

We use the following notations
\begin{eqnarray}
M=\sum_k m_k,\\
I=\sum_k m_k q_k^2=M^{-1}\sum_{i<j}m_i m_j (q_i-q_j)^2,\\
K=\sum_k m_k v_k^2,\\
L=\sum_k q_k \wedge p_k,
\end{eqnarray}
and take the centre of mass to be the origin
\begin{equation}
\sum_k m_k q_k =0.
\end{equation}

Now let us prove the equation 
\eref{eqDefOfRatio0}.
Let $\kappa (t)$ be the ratio of magnification
\begin{equation}
\label{eqDefOfRatio}
\kappa (t)=\frac{|p_k(t)|}{|q_i(t)-q_j(t)|}.
\end{equation}
Here $(i, j, k)=(1,2,3), (2,3,1), (3,1,2)$,
and 
we always use this convention 
when indexes $i, j, k$ appear in one equation.
Then we get
\begin{eqnarray}
\label{eqRatio}
\frac{\kappa^2}{m_1 m_2 m_3 }
=\frac{p_k^2/m_k}{m_i m_j(q_i-q_j)^2}
=\frac{\sum_k p_k^2/m_k}{\sum_{i<j} m_i m_j(q_i-q_j)^2}
=\frac{K}{M I},
\end{eqnarray}
%
%
which leads to
\begin{equation}
\kappa=\sqrt{
		\frac{m_1 m_2 m_3 K}{M I}
	},
\end{equation}
i.e., 
the equation \eref{eqDefOfRatio0}
and
\begin{equation}
\label{eqPerimeterQV1}
\frac{m_i m_j (q_i-q_j)^2}{M I}
=\frac{p_k^2/m_k}{K}
=\frac{m_k v_k^2}{K}.
\end{equation}

Now,
consider the oriented area of 
the ``synchronised similar triangles''.
Since the two triangles are inversely similar,
we get the equation \eref{eqArea0} as follows
\begin{eqnarray}
p_1 \wedge p_2
=-\kappa^2 (q_2-q_1)\wedge (q_3-q_1)
\nonumber\\
=-\frac{K}{M I} m_1 m_2 m_3
	(q_1 \wedge q_2+q_2\wedge q_3+q_3\wedge q_1)
\nonumber\\
=-\frac{K}{I} m_1 m_2 q_1 \wedge q_2,
\end{eqnarray}
%
where we have used
the relation of 
$
m_1 m_2 q_1 \wedge q_2
=m_2 m_3 q_2\wedge q_3
=m_3 m_1q_3\wedge q_1,
$
which follows from 
$\sum_i m_i q_i=0$.

Note that
the following identity holds
for vectors $\eta_i$ 
which satisfy $\sum_i m_i \eta_i =0$,
\begin{equation}
\label{eqIdentityI}
m_i m_j (\eta_i-\eta_j)^2
+
M m_k \eta_k^2
=(m_i+m_j)\sum_\ell m_\ell \eta_\ell^2.
\end{equation}
Using this identity,
the equation \eref{eqPerimeterQV1}
can be written as
\begin{equation}
\frac{m_i m_j (v_i-v_j)^2}{M K}=\frac{m_k q_k^2}{I}. \label{eqPerimeterQV2}
\end{equation}
Then,
the equations \eref{eqPerimeterQV1},
\eref{eqPerimeterQV2}
and 
\eref{eqIdentityI}
yield the following interesting equation
\begin{equation}
\frac{m_k q_k^2}{I}+\frac{m_k v_k^2}{K}
=\frac{m_i m_j (q_i-q_j)^2}{M I}+\frac{m_i m_j (v_i-v_j)^2}{M K}
=\frac{m_i+m_j}{M}. \label{eqPerimeterQV3}
\end{equation}
The equation \eref{eqPerimeterQV2}
shows that
\begin{equation}
\frac{m_k |q_k|}{|v_i - v_j|}
=
\sqrt{
	\frac{m_1m_2m_3 I}{MK}
}.
\end{equation}
Therefore,
with the equation \eref{eqArea0},
we conclude that
the triangle whose vertexes are $v_i$
and the triangle whose perimeters are $m_k q_k$,
are always inversely similar.
Thus,
the role of $q_i$ and $v_i$
are completely equivalent.
Indeed,
the following 
purely algebraic theorem
holds.
\begin{theorem}
\label{algebraic}
Consider two triplets of two dimensional vectors
$\{\xi_i\}$, $\{\bar{\xi}_i\}$ and a triplet of scalars $\{\mu_i\}$,
$i=1,2,3$,
which satisfy
\begin{equation}
\sum_i \mu_i \xi_i=0,\;
\sum_i \mu_i \bar{\xi}_i=0,\;
\sum_i \mu_i \xi_i \wedge \bar{\xi}_i=0,\;
\sum_i \mu_i \xi_i \cdot \bar{\xi}_i=0.
\end{equation}
Let $I$ be the ``moment function'' defined by
\begin{equation}
I(\eta)=\sum_i \mu_i \eta_i^2.
\end{equation}
Then,
we have the following
three equivalent equations
\begin{eqnarray}
\frac{\mu_k \xi_k^2}{I(\xi)}
=\frac{\mu_i \mu_j (\bar{\xi}_i-\bar{\xi}_j)^2}{(\mu_1+\mu_2+\mu_3) I(\bar{\xi})},
\label{eqSimilarXi1}\\
\frac{\mu_i \mu_j (\xi_i-\xi_j)^2}{(\mu_1+\mu_2+\mu_3) I(\xi)}
=\frac{\mu_k \bar{\xi}_k^2}{I(\bar{\xi})},
\label{eqSimilarXi2}\\
\fl
\frac{\mu_k \xi_k^2}{I(\xi)}+\frac{\mu_k \bar{\xi}_k^2}{I(\bar{\xi})}
=\frac{\mu_i \mu_j ({\xi}_i-{\xi}_j)^2}{(\mu_1+\mu_2+\mu_3) I({\xi})}
+\frac{\mu_i \mu_j (\bar{\xi}_i-\bar{\xi}_j)^2}{(\mu_1+\mu_2+\mu_3) I(\bar{\xi})}
=\frac{\mu_i+\mu_j}{\mu_1+\mu_2+\mu_3}
\label{eqSimilarXi3}
\end{eqnarray}
and 
\begin{equation}
\label{eqAreaXi}
\frac{\xi_i\wedge \xi_j}{I(\xi)}
+
\frac{\bar{\xi}_i\wedge \bar{\xi}_j}{I(\bar{\xi})}
=0.
\end{equation}
\end{theorem}

\textbf{Remark}
Therefore,
triangle whose vertexes are $\xi_i$, and 
triangle whose perimeters are $\mu_i \bar{\xi}$,
are inversely similar. 
Equivalently,
triangle whose vertexes are $\bar{\xi}$, and 
triangle whose perimeters are $\mu_i \xi_i$,
are also inversely similar.


\textbf{Proof:}
Let us 
denote
the Jacobi coordinates in $\xi$~space
and $\bar{\xi}$ space 
respectively by
$\{a,b\}$ and $\{\bar{a},\bar{b}\}$,
where
\begin{eqnarray}
a
	&=&\rho\; \frac{\mu_1 \xi_1+\mu_2 \xi_2}{\mu_1+\mu_2}
	=-\rho\; \frac{\mu_3 \xi_3}{\mu_1+\mu_2},\\
b
	&=&\sigma\; (\xi_1-\xi_2)
\end{eqnarray}
with
similar expressions for the ``bar'' variables,
and
\begin{equation}
\rho=\sqrt{
		\frac{(\mu_1+\mu_2)(\mu_1+\mu_2+\mu_3)}{\mu_3}
	},
\;
\sigma=\sqrt{
		\frac{\mu_1 \mu_2}{\mu_1+\mu_2}
	}.
\end{equation}
The inverse relations are
\begin{eqnarray}
\xi_1 = \frac{a}{\rho}+\frac{\mu_2}{\mu_1+\mu_2}\;\frac{b}{\sigma},\\
\xi_2 = \frac{a}{\rho}-\frac{\mu_1}{\mu_1+\mu_2}\;\frac{b}{\sigma},\\
\xi_3 = - \frac{\mu_1+\mu_2}{\mu_3}\;\frac{a}{\rho}
\end{eqnarray}
and similar expressions for the ``bar'' variables.
Then, the equations
$\sum_i \mu_i \xi_i=0$,
$\sum_i \mu_i \bar{\xi}_i=0$
are automatically satisfied
and the equations
$\sum_i \mu_i \xi_i \wedge \bar{\xi}_i =0$,
$\sum_i \mu_i \xi_i \cdot \bar{\xi}_i =0$
yield
\begin{eqnarray}
a\wedge \bar{a} + b \wedge \bar{b}=0,\label{eqOuterAB}\\
a\cdot \bar{a} + b \cdot \bar{b}=0.\label{eqInnerAB}
\end{eqnarray}

Let us use the polar coordinate for 
the vectors $a, b, \bar{a}, \bar{b}$:
\begin{equation}
a=(|a|, \alpha),
b=(|b|, \beta)
\end{equation}
and similar notations for the ``bar'' variables.
%
The equations \eref{eqOuterAB} and \eref{eqInnerAB}
then
yield
\begin{eqnarray}
|a|\;|\bar{a}|\sin(\bar{\alpha}-\alpha)
+
|b|\;|\bar{b}|\sin(\bar{\beta}-\beta)
=0,\\
|a|\;|\bar{a}|\cos(\bar{\alpha}-\alpha)
+
|b|\;|\bar{b}|\cos(\bar{\beta}-\beta)
=0,
\end{eqnarray}
%
which give
\begin{eqnarray}
a^2\;\bar{a}^2=b^2\;\bar{b}^2,\label{eqMagnitude}\\
\bar{\beta}-\beta=\bar{\alpha}-\alpha+\pi.\label{eqAngle}
\end{eqnarray}

From the equation \eref{eqMagnitude}
we get
\begin{equation}
\label{eqRatioOfMagnitude}
\frac{\bar{b}^2}{a^2}
=\frac{\bar{a}^2}{{b}^2}
=\frac{\bar{a}^2+\bar{b}^2}{a^2+b^2},
\end{equation}
where
the second equality is a consequence 
of the first equality.
Then, we have the following three equivalent equations
\begin{eqnarray}
\frac{a^2}{a^2+b^2}=\frac{\bar{b}^2}{\bar{a}^2+\bar{b}^2},\label{eqSimilarA1}\\
\frac{b^2}{a^2+b^2}=\frac{\bar{a}^2}{\bar{a}^2+\bar{b}^2},\label{eqSimilarA2}\\
\frac{a^2}{a^2+b^2}+\frac{\bar{a}^2}{\bar{a}^2+\bar{b}^2}
=\frac{b^2}{a^2+b^2}+\frac{\bar{b}^2}{\bar{a}^2+\bar{b}^2}=1.\label{eqSimilarA3}
\end{eqnarray}
Note that  
$
a^2+b^2
=\sum_i \mu_i \xi_i^2
=I(\xi)
$
is the ``moment function''
defined in the theorem~\ref{algebraic}.
Rewriting the equations 
\eref{eqSimilarA1}--\eref{eqSimilarA3}
by $\xi$ and $\bar{\xi}$ variables,
we get
the equations
\eref{eqSimilarXi1}--\eref{eqSimilarXi3}
in the theorem \ref{algebraic}.

From the equation
\eref{eqAngle}
we get
\begin{equation}
\bar{\beta}-\bar{\alpha}
=\beta-\alpha+\pi.
\end{equation}
Moreover,
the 
product of
equations
\eref{eqSimilarA1} and \eref{eqSimilarA2}
yield
\begin{equation}
\label{eqRatioOfMagnitude2}
\frac{|a|\;|b|}{a^2+b^2}
=
\frac{|\bar{a}|\;|\bar{b}|}{\bar{a}^2+\bar{b}^2}.
\end{equation}
Therefore,
the following equation is obvious
\begin{equation}
\frac{a \wedge b}{a^2+b^2}
+
\frac{\bar{a} \wedge \bar{b}}{\bar{a}^2+\bar{b}^2}
=
\frac{|a|\;|b| \sin(\beta-\alpha)}{a^2+b^2}
+
\frac{|\bar{a}|\;|\bar{b}|\sin(\bar{\beta}-\bar{\alpha})}{\bar{a}^2+\bar{b}^2}
=0.
\end{equation}
Rewriting this equation by $\xi$ and $\bar{\xi}$,
we finally get
the equation \eref{eqAreaXi}
in the theorem \ref{algebraic}.

\section{Constant along the orbit under homogeneous potentials}
\label{sec_homopot}
Let us
consider the orbit
with zero angular momentum
and constant moment of inertia
under the 
potential
$V_\alpha$
defined 
by
the equation \eref{eqDefOfV}.

Since $I=\mbox{constant}$,
the Jacobi-Lagrange identity
yields
\begin{equation}
\label{eqJacobiLagrange}
0=\frac{d^2 I}{dt^2}
=\cases{
	2(K-\alpha V_\alpha)
	=2
		\left(
			K-\sum_{i<j}m_i m_j r_{ij}^{\alpha}
		\right)
		& for $\alpha \ne 0$\\
	2
		\left(
			K - \sum_{i<j} m_i m_j
		\right)
		& for $\alpha=0$.
	}
\end{equation}
Therefore
for $\alpha \ne -2$, the
constant moment of inertia
along the orbit
yields
\begin{equation}
\label{kineticEnergy}
K
=\cases{
	\sum_{i<j}m_i m_j r_{ij}^{\alpha}
	=\frac{\alpha}{2+\alpha}E
		& for $\alpha \ne 0,-2$\\
	\sum_{i<j} m_i m_j
	=2
		\left(
			E-\sum_{ij}m_i m_j \log r_{ij}
		\right)
		& for $\alpha=0$,
	}
\end{equation}
%
and
both
the kinetic energy $K/2$ and the potential energy $V_\alpha$ 
are
constant. 
As a consequence of this equation, $r_{ij}$ cannot be zero 
if $\alpha \le 0$ and $\alpha \ne -2$. 
That is, there is no collision along the orbit.
For $\alpha = -2$,
on the other hand,
$K$ and $V_{-2}$ can vary along the orbit
keeping the energy balance
\begin{equation}
  \label{kineticEnergyMinus2}
  K=-2V_{-2}=\sum_{i<j} \frac{m_i m_j}{r_{ij}^2}
  \mbox{ for } \alpha=-2,
\end{equation}
with
zero total energy.

Moreover,
if the angular momentum is zero,
then 
by the use of
the equation \eref{eqDefOfRatio0},
$r_{ij}^2$ can be expressed by the momentum $p_k^2$,
\begin{equation}
r_{ij}^2
=\frac{MI}{m_1m_2m_3K}\; p_k^2.
\end{equation}
%
Then the equations \eref{kineticEnergy}
or \eref{kineticEnergyMinus2}
yields
\begin{equation}
\label{eqConstantAlongTheOrbit}
\cases{
\sum_{(ijk)} m_i m_j |p_k|^\alpha
=K
	\left(
		\frac{m_1 m_2 m_3 K}{MI}
	\right)^{\frac{\alpha}{2}}
	&for $\alpha\ne 0$\\
\sum_{(ijk)} m_i m_j \log |p_k|
=E+\frac{K}{2}\log \frac{m_1m_2m_3K}{MI}
	&for $\alpha= 0$.
}
\end{equation}
%
The right-hand sides
of the above equations are
constant for all $\alpha$.
Note that
for $\alpha = -2$,
\begin{equation}
  \label{eqNewConstant}
  {\rm r.h.s.}
  =\frac{M I}{m_1 m_2 m_3},
\end{equation}
%
is also
constant along the orbit.
Thus we get the equation
\eref{eqConstant0}.
As a consequence of this equation,
$p_k$ cannot be zero if $\alpha \le 0$,
meaning that
the bodies cannot stop.

For $\alpha \ne -2$, 
the orbit of 
$L=0$ and $I=\mbox{constant}$
is strictly constrained
by
the condition for $K$ and $V_{\alpha}$ to be constant
and
the condition \eref{eqConstantAlongTheOrbit}.
For $\alpha=-2$,
on the other hand,
the condition
\eref{kineticEnergyMinus2} 
is relatively loose
%
and actually the figure-eight orbit
shown in \fref{figCtCnCo}
is that with  $L=0$ and $I=\mbox{constant}$.

\section{Similarity in the momentum space and the force space}
\label{sec_force_space}
The similarity for
the momentum space and 
the force space
also holds.
This is interesting,
but this similarity does not produce any new information.

Differentiate the equation \eref{eqConstantAlongTheOrbit} with 
respect to $t$,
we get
\begin{equation}
m_1 m_2 |p_3|^{\alpha-2}\; p_3 \cdot f_3
+m_2 m_3 |p_1|^{\alpha-2}\;p_1 \cdot f_1
+m_3 m_1 |p_2|^{\alpha-2}\;p_2 \cdot f_2
=0,
\end{equation}
where
\begin{equation}
f_k = \frac{dp_k}{dt}
\end{equation}
represents the force acting on the body $k$.
Substituting the equation \eref{eqDefOfRatio0}
into the above equation, we get
\begin{equation}
\label{eqPFInner}
m_1 m_2 r_{12}^{\alpha-2}\; p_3 \cdot f_3
+m_2 m_3 r_{23}^{\alpha-2}\;p_1 \cdot f_1
+m_3 m_1 r_{31}^{\alpha-2}\;p_2 \cdot f_2
=0.
\end{equation}
On the other hand, we have the following equality
\begin{equation}
\label{eqPFOuter}
m_1 m_2 r_{12}^{\alpha-2}\; p_3 \wedge f_3
+m_2 m_3 r_{23}^{\alpha-2}\;p_1 \wedge f_1
+m_3 m_1 r_{31}^{\alpha-2}\;p_2 \wedge f_2
=0,
\end{equation}
%
which is proved by
substituting
the equations of motion into 
the forces $f_k$
and showing that l.h.s. becomes
\begin{equation}
\fl
m_1 m_2 m_3 
	\left(
		\sum_k q_k \wedge p_k
	\right)
	\left(
		m_1 r_{31}^{\alpha-2} r_{12}^{\alpha-2}
		+
		m_2 r_{12}^{\alpha-2} r_{23}^{\alpha-2}
		+
		m_3 r_{23}^{\alpha-2} r_{31}^{\alpha-2}
	\right)
=0.
\end{equation}

Let
\begin{equation}
\mu_1^{-1}=m_2 m_3 r_{23}^{\alpha-2},\;
\mu_2^{-1}=m_3 m_1 r_{31}^{\alpha-2},\;
\mu_3^{-1}=m_1 m_2 r_{12}^{\alpha-2}
\end{equation}
and
\begin{equation}
\xi_i=\frac{p_i}{\mu_i},\;
\bar{\xi}_i=\frac{f_i}{\mu_i}.
\end{equation}
Then,
the equations
$\sum p_i=0$,
$\sum f_i=0$,
 \eref{eqPFInner} and \eref{eqPFOuter} are
rewritten as
\begin{equation}
\sum_i \mu_i \xi_i =0,\;\;
\sum_i \mu_i \bar{\xi}_i =0,\;\;
\sum_i \mu_i \xi_i \cdot \bar{\xi}_i =0,\;\;
\sum_i \mu_i \xi_i \wedge \bar{\xi}_i =0.
\end{equation}
By the theorem \ref{algebraic},
the triangle 
whose vertexes are
$\bar{\xi}_k
=f_k/\mu_k
=m_1 m_2 m_3 r_{ij}^{\alpha-2} d^2 q_k/dt^2
$,
and
the triangle
whose perimeters are
$\mu_i \xi_i=p_i$,
are the
``synchronised similar triangles''.
They are always inversely similar.
%
However, this similarity
gives no new information,
because they are equivalent to 
the similarity in 
$q{-}v$ variables.
%
%

\section{Synchronised similar triangles for $L=0$ orbit}
\label{sec_triangle_gen}
In this section,
we consider 
general three-body orbits
with $L=0$,
but not with the assumption of $I=\mbox{constant}$.
Even in this case,
we can find 
the
``synchronised similar triangles''.
Consider the scaled position and
the velocity of the scaled position
defined by the equations
\eref{eq_scaledq}
and
\eref{eq_scaledv}.
%
We can easily verify the following equalities for the scaled variables,
\begin{eqnarray}
  \sum_i m_i \tilde{q}_i=0,\;
  \sum_i m_i \tilde{v}_i=0,\;
  \sum_i m_i \tilde{q}_i \wedge \tilde{v}_i =0,\;
  \sum_i m_i \tilde{q}_i \cdot \tilde{v}_i=0.
  \label{eqJeq1}
\end{eqnarray}
%
By the theorem \ref{algebraic},
triangle whose vertexes are $\tilde{q}_i$,
and 
triangle whose perimeters are $m_i \tilde{v}_i$,
are the
``synchronised similar triangles''.
Therefore, 
all the equalities in the section \ref{basicProperty}
hold for the variables
$\tilde{q}_i$ and $\tilde{v}_i$.
%
Rewriting the equality for 
$\tilde{q}_i$ and $\tilde{v}_i$
by the original variables
$q_i$ and $v_i$,
we 
have useful equalities
for $L=0$ and $I \ne \mbox{constant}$ orbits.
For example, the equality \eref{eqArea0}
for the scaled variables
\begin{equation}
\label{eqAreaScaled}
\tilde{q}_i \wedge \tilde{q}_j
+
\frac{\tilde{v}_i \wedge \tilde{v}_j}
	{\sum_k m_k \tilde{v}_k^2}
=0
\end{equation}
yields
the equality \eref{eq_AreaGeneral},
since
\begin{eqnarray}
\tilde{q}_i \wedge \tilde{q}_j
=\frac{q_i \wedge q_j}{I},\\
\tilde{v}_i \wedge \tilde{v}_j
=
\frac{1}{I}
	\left(
		v_i \wedge v_j
		+
		q_i \wedge q_j \;
		\frac{1}{4I^2}
			\left(
				\frac{dI}{dt}
			\right)^2
		-
		\frac{1}{2I}
		\frac{dI}{dt}
		\frac{d}{dt}\left( q_i \wedge q_j \right)
	\right),\\
\sum_k m_k \tilde{v}_k^2
=
\frac{K}{I}
-
\frac{1}{4I^2}
	\left(
		\frac{dI}{dt}
	\right)^2.
\end{eqnarray}

\section{Infinitely many syzygies or collisions for $L=0$ orbit}
\label{sec_syzygies}
In this section,
we consider  $L=0$ orbit
under  the potential energy $V_\alpha$.
We do not assume $I=\mbox{constant}$.
We derive an equation
of motion
for the oriented area
defined by the positions $q_i$ of three bodies,
%
and
prove that
any 
three-body orbits
with $L=0$
under the potential $V_\alpha$
with $\alpha \le 2$
have infinitely many syzygies
or collisions.

Let 
$
\Delta=2^{-1} (q_2-q_1)\wedge (q_3-q_1)
$
be the oriented area
and 
\begin{equation}
\Lambda_{ij}=q_i \wedge q_j
\end{equation}
be twice of 
the oriented area of the triangle 
defined by $q_i$, $q_j$ and the origin. 
Note that
\begin{equation}
\Delta
=\frac{1}{2}
	\left(
		\Lambda_{12}+\Lambda_{23}+\Lambda_{31}
	\right).
\end{equation}
The second derivative of $\Lambda_{ij}$ with respect to time $t$
is
\begin{equation}
\label{eqd2Lambda}
\frac{d^2 \Lambda_{ij}}{dt^2}
=2 \; v_i \wedge v_j
	+ \frac{d^2 q_i}{dt^2}\wedge q_j
	+q_i \wedge \frac{d^2 q_j}{dt^2}.
\end{equation}
By virtue of \eref{eq_AreaGeneral},
we have
\begin{equation}
\label{eqAreavv}
v_i \wedge v_j
=
-\frac{K}{I} \Lambda_{ij}
+
\frac{1}{2I}
\frac{dI}{dt}
\frac{d \Lambda_{ij}}{dt}.
\end{equation}
Using the equation of motion for $q_i$,
we get
\begin{equation}
\label{eqAread2qq}
\frac{d^2 q_i}{dt^2}\wedge q_j
	+q_i \wedge \frac{d^2 q_j}{dt^2}
=
-\Lambda_{ij}
\sum_{k\ell} (m_k+m_\ell)r_{k\ell}^{\alpha-2}.
\end{equation}
Substituting the equations 
\eref{eqAreavv} and \eref{eqAread2qq}
into
the equation
\eref{eqd2Lambda},
we get the equation 
for $\Lambda_{ij}$
\begin{equation}
\frac{d^2 \Lambda_{ij}}{dt^2}
=
-
	\left(
		\frac{2K}{I}
		+
		\sum_{k\ell} (m_k+m_\ell)r_{k\ell}^{\alpha-2}
	\right)
	\Lambda_{ij}
+
\frac{1}{I}
\frac{dI}{dt}
\frac{d \Lambda_{ij}}{dt},
\end{equation}
and the 
equation for $\Delta$
of the same form,
\begin{equation}
\label{eqBasicForDelta}
\frac{d^2 \Delta}{dt^2}
=
-
	\left(
		\frac{2K}{I}
		+
		\sum_{k\ell} (m_k+m_\ell)r_{k\ell}^{\alpha-2}
	\right)
	\Delta
+
\frac{1}{I}
\frac{dI}{dt}
\frac{d \Delta}{dt},
\end{equation}
%
which is equivalent to the
equation \eref{eqForDelta}.

Now,
let us prove that
the function $\Delta(t)$
or equivalently
\begin{equation}
	S(t)=\frac{\Delta(t)}{\sqrt{I(t)}}
\end{equation} 
has
infinitely many zeros
for $\alpha \le 2$.
Note that the zeros of $\Delta(t)$
and $S(t)$ are one to one.
There are three cases
when they take zero:
syzygy without collision,
two-body collision
and triple collision $I \to 0$.
%
%
Thus, infinitely many zeros
of $S(t)$ correspond to
infinitely many syzygies
or collisions
of the orbit.
Montgomery proved this property 
making use of
the equation of motion for $\Delta/I$.
We give here 
another
proof from a different point of view.
%
Our proof consists of three steps.

The first step:
Eliminate the first derivative term $d\Delta/dt$
in the equation \eref{eqBasicForDelta}.
To do this,
we consider the equation for $S(t)$
instead of $\Delta(t)$.
Then, the equation \eref{eqBasicForDelta} is
equivalent to
\begin{eqnarray}
\label{eqForS}
\frac{d^2 S}{dt^2}
&=&
-
\left\{
	\sum_{i<j} (m_i+m_j)r_{ij}^{\alpha-2}
	+
	\frac{2K}{I}
	+
	\frac{1}{2I}
	\frac{d^2 I}{dt^2}
	-
	\frac{3}{4I^2}
	\left(
		\frac{dI}{dt}
	\right)^2
\right\}
S
\nonumber\\
&=&
-
\left\{
	\frac{M}{I}\sum_{(ijk)} m_k q_k^2 r_{ij}^{\alpha-2}
	+
	\frac{3K}{I}
	-
	\frac{3}{4I^2}
	\left(
		\frac{dI}{dt}
	\right)^2
\right\}
S
,
\end{eqnarray}
%
where
we have used two identities
$d^2 I/dt^2=2(K-\alpha V_\alpha)$ and
\eref{eqIdentityI} for $m_i + m_j$.
%

The second step:
Write the equation \eref{eqForS} as
\begin{eqnarray}
\frac{d^2 S}{dt^2}=-\omega^2 S,\\
\omega^2=
	\frac{M}{I}\sum_{(ijk)} \frac{m_k q_k^2}{r_{ij}^{2-\alpha}}
	+
	\frac{3K}{I}
	-
	\frac{3}{4I^2}
	\left(
		\frac{dI}{dt}
	\right)^2,
\end{eqnarray}
and
show that $\omega^2$ is bounded from below
by a positive constant
 $\omega^2 \ge \omega_0^2>0$
for $\alpha \le 2$.
The following inequalities hold
\begin{eqnarray}
\left(
	\frac{dI}{dt}
\right)^2
=
\left(
	2\sum_i m_i q_i v_i
\right)^2
\le
4
\left(
	\sum_i m_i q_i^2
\right)
\left(
	\sum_i m_i v_i^2
\right)
=
4IK,
\\
\sum_{(ijk)}
	\frac{m_k q_k^2}{r_{ij}^{2-\alpha}}
\ge
\sum_{(ijk)}
	m_k q_k^2
	\left(
		\frac{m_\textrm{min}^2}{MI}
	\right)^{(2-\alpha)/2}
=
I 
	\left(
		\frac{m_\textrm{min}^2}{MI}
	\right)^{(2-\alpha)/2},
\end{eqnarray}
%
where
we have used
an inequality that
$MI=\sum_{i<j}m_i m_j r_{ij}^2
\ge 
m_i m_j r_{ij}^2
\ge
m_{\textrm{min}}^2 r_{ij}^2
$
and 
$2 \ge \alpha$.
The symbol 
$m_{\textrm{min}}$
represents 
the minimum value of 
$\{m_k\}$.
%
%
Then,
we get the following inequality
\begin{equation}
\omega^2
\ge
M
\left(
	\frac{m_{\textrm{min}}^2}{M I}
\right)^{(2-\alpha)/2}
\ge
M
\left(
	\frac{m_{\textrm{min}}^2}{MI_{\textrm{max}}}
\right)^{(2-\alpha)/2}
=
\omega_0^2
>0
\end{equation}
where
we have used the fact that
the orbit is bounded
$I \le I_{\textrm{max}}$.

The last step:
Since the 
restoring
force for $S(t)$ is always stronger than
that 
of 
a harmonic oscillator
with $\omega=\omega_0$,
it is natural to expect that
$S(t)$ has infinitely many zeros
and intervals of zeros are shorter than $T_0=\pi/\omega_0$.
To prove this 
we show that
for
any initial conditions $S(0)$ and $dS/dt\;(0)$,
the function $S(t)$ 
vanishes
before $t=T_0$.
Let us consider a harmonic oscillator $A(t)$
which has the period $2T_0=2\pi/\omega_0$
and satisfies the same initial conditions as $S(t)$,
\begin{eqnarray}
\frac{d^2 A(t)}{dt^2}=-\omega_0^2\; A(t),\\
A(0)=S(0),\;\;
\frac{dA}{dt}(0)=\frac{d{S}}{dt}(0).
\end{eqnarray}
Define a function $Z(t)$ by
\begin{equation}
Z(t)
=S(t)\frac{d A(t)}{dt}
	-\frac{d {S}(t)}{dt}A(t).
\end{equation}
Then, the first derivative with respect to $t$ yields
\begin{equation}
\label{eqdZ}
\frac{dZ(t)}{dt}
=S(t)\frac{d^2 A(t)}{dt^2}
	-\frac{d^2 {S}(t)}{dt^2}A(t)
	=	\left(
		\omega^2 -\omega_0^2
	\right)
	S(t) A(t).
\end{equation}
Without loss of generality,
we can take $A(0)=S(0)>0$.
Let $t_0>0$ be the first time when $A(t)$ 
vanishes.
Then we have $t_0<T_0$ and $dA/dt(t_0)<0$.
%
To derive a contradiction, suppose that 
$S(t)>0$ for $0\le t \le t_0$.
Then, 
$\left(
		\omega^2 -\omega_0^2
\right)
S(t) A(t) \ge 0$
for $0\le t \le t_0$.
%
Since
$Z(0)=0$ and $A(t_0)=0$,
integration of the equation \eref{eqdZ} yields
\begin{equation}
Z(t_0)
=S(t_0)\frac{dA}{dt}(t_0)
=\int_0^{t_0}\left(
		\omega^2 -\omega_0^2
	\right)
	S(s) A(s)
	ds
\ge 0.
\end{equation}
Since $dA(t)/dt(t_0)<0$,
we get
$S(t_0) \le 0$,
%
which
is a contradiction.
Therefore,
$S(t)$ must have zero before $t = t_0 < T_0$.

\section{Final remarks}
\label{sec_final}
This work 
may
give  rich information
for 
some related problems.
%
The first one is
to investigate the nature of 
the figure-eight solution under $-1/r_{ij}^2$ potential
which
has 
zero angular momentum
and
constant moment of inertia.
We hope that
the equalities
\eref{eqDefOfRatio0}, \eref{eqArea0} and \eref{eqConstant0}
would be useful for further understanding
of the figure-eight solution under this potential.

The second problem is
to find  a conceptual proof of
Chenciner's problem 13
in his lecture at Taiyuan~\cite{chenq13}.
Consider the figure-eight solution
under the homogeneous 
or logarithmic potential energy
$V_\alpha$.
Chenciner's problem is the following:
{\it
Show that the moment of inertia of the ``Eight''
stays constant only when $\alpha=-2$.
}
In the equal-masses three-body problem,
FFO~\cite{ffoI} showed that
a motion 
satisfying the following
three conditions
exists 
under the potentials $V_\alpha$
if and only if $\alpha=-2, 2, 4$.
The conditions are
(i) $L=0$,
(ii) $I=\mbox{constant}$
and 
(iii) one body passes through 
the centre of mass.
Then FFO explicitly proved that
the orbits for $\alpha=2, 4$
are not the figure-eight.
%
This solves 
Chenciner's problem.
But the method of FFO
was a kind of ``brute-force'' one,
in which
they calculated the derivative
of the moment of inertia
with respect to time to the eighth order
explicitly.
A more ``conceptual'' proof
would be appreciated.
%
The present work shows
what will happen
if $L=0$ and $I=\mbox{constant}$
orbit exist for $V_\alpha$ with $\alpha \ne -2$.
%
As discussed in the section \ref{sec_homopot},
these orbits are strictly constrained.

The third problem is
to understand the nature of
three-body orbits
with zero angular momentum
under
various potential energy
$V_\alpha$.
We expect that
the information for the scaled variables
will be useful
for this problem.
%
%

\ack
The authors would like to thank
RIMS 
(Research Institute for Mathematical Sciences)
for funding a workshop 
`` Developments and Applications of Dynamical Systems Theory''
where the authors were motivated to start this work.
One of the author (T.F.)
would like to express 
his
special thanks to
Richard Montgomery
and Haruo Yoshida
for their
stimulating discussions.


\end{document}